# Automated Machine Learning in the smart construction era: Significance and accessibility for industrial classification and regression tasks

Rui Zhao[1]*, Zhongze Yang[2], Dong Liang[3], Fan Xue[4]

**Abstract:** This paper explores the application of automated machine learning (AutoML) techniques to the construction industry, a sector vital to the global economy. Traditional ML model construction methods were complex, time-consuming, reliant on data science expertise, and expensive. AutoML shows the potential to automate many tasks in ML construction and to create outperformed ML models. This paper aims to verify the feasibility of applying AutoML to industrial datasets for the smart construction domain, with a specific case study demonstrating its effectiveness. Two data challenges that were unique to industrial construction datasets are focused on, in addition to the normal steps of dataset preparation, model training, and evaluation. A real-world application case of construction project type prediction is provided to illustrate the accessibility of AutoML. By leveraging AutoML, construction professionals without data science expertise can now utilize software to process industrial data into ML models that assist in project management. The findings in this paper may bridge the gap between data-intensive smart construction practices and the emerging field of AutoML, encouraging its adoption for improved decision-making, project outcomes, and efficiency.

[1] Rui Zhao
Department of Real Estate and Construction, The University of Hong Kong, Hong Kong, China
E-mail: jeremyrui.zhao@connect.hku.hk

*: Corresponding author

[2] Zhongze Yang
Department of Real Estate and Construction, The University of Hong Kong, Hong Kong, China
E-mail: zoeyang@connect.hku.hk

[3] Dong Liang
Department of Real Estate and Construction, The University of Hong Kong, Hong Kong, China
E-mail: leodong@hku.hk

[4] Xue Fan
Department of Real Estate and Construction, The University of Hong Kong, Hong Kong, China
E-mail: xuef@hku.hk

# 1 Introduction

The construction industry plays a vital role in the global economy, and the ability to make informed decisions is crucial for successful project execution[1]. Smart construction is an evolution within the construction industry that leverages digital technology and advanced building practices to improve efficiency, safety, and sustainability. Numerous researchers are also incorporating computer technologies such as blockchain, machine learning, and digital twins into smart construction[2-4]. Machine learning (ML) plays a key role in the smart construction domain. Examples of ML are supply chain anagement, quality control, construction site safety, project management, and scheduling[5].

Traditionally, constructing predictive ML models in the construction domain has been complex and time-consuming. It requires expertise in data preprocessing, feature engineering, algorithm selection, hyperparameter tuning, and model evaluation. With the advent of automated machine learning (AutoML) techniques, there is an opportunity to revolutionize how construction datasets are analyzed and utilized. AutoML offers the potential to streamline and automate the process of developing machine learning models, making it accessible to professionals without extensive data science expertise[6]. With AutoML, many of these tasks can be automated, significantly reducing the manual effort required[7-8].

This paper aims to test the feasibility and accessibility of applying AutoML techniques for smart construction. First, six industrial datasets from different construction fields are employed to validate the robustness and versatility of AutoML. Furthermore, the datasets focus on several challenges, e.g., limited data samples and imbalanced classes, which are common in construction industrial data. For accessibility, both commercial AutoML software and open-sourced Python libraries are involved and tested.

By leveraging AutoML, construction professionals can leverage the wealth of data available in the industry, including project plans, sensor data, labor records, and more. This data can be used to generate predictive models that can assist in various aspects of construction project management, such as predicting project types, estimating resource requirements, identifying potential risks, and optimizing project timelines[9].

This paper aims to bridge the gap between traditional construction practices and the emerging field of automated machine learning. By showcasing the feasibility and benefits of AutoML in construction datasets, we aim to encourage the adoption of these techniques in the industry, enabling more informed decision-making, improved project outcomes, and increased efficiency.

# 2 Related works

## 2.1 Machine Learning in the construction industry

Machine learning has emerged as a valuable tool in the construction industry, offering new data analysis and decision-making possibilities. Predictive analytics is a prominent application of machine learning, where models trained on historical project data can forecast outcomes, estimate durations, predict resource requirements, and identify potential risks and delays[10]. These predictive capabilities enable construction professionals to optimize resource allocation, schedule projects effectively, and proactively manage risks.

Another significant application of machine learning in construction is anomaly detection. Machine learning algorithms can detect patterns and identify anomalies such as safety incidents, equipment failures, and material shortages by analyzing sensor data, construction site images, and other relevant sources. Early detection of anomalies enables prompt interventions, reducing the likelihood of accidents and mitigating risks[11].

Moreover, machine learning contributes to process optimization within the construction industry. Machine learning algorithms identify patterns and correlations that improve process efficiency by analyzing diverse datasets encompassing project designs, material properties, labor productivity, and equipment performance, which includes optimizing resource allocation, identifying bottlenecks, and suggesting alternative construction methods or materials to enhance productivity and reduce costs[12].

## 2.2 Automated machine learning

Machine learning techniques have demonstrated significant potential in the construction industry but are not without limitations. One of the primary challenges is the complexity and expertise required in the manual development and fine-tuning of machine learning models[13]. This process involves various tasks, such as data preprocessing, feature engineering, algorithm selection, and hyperparameter tuning, which can be time-consuming and require specialized knowledge[14]. Additionally, the construction domain often deals with heterogeneous and complex datasets, making it difficult to navigate and extract meaningful insights [10] manually.

To address these limitations and streamline the machine learning process, AutoML has emerged as a promising solution. AutoML aims to automate many manual tasks involved in model development, reducing the dependency on human expertise and time-intensive processes[15]. By automating steps such as feature engineering, algorithm selection, and hyperparameter optimization, AutoML empowers construction professionals with limited data science backgrounds to leverage machine learning effectively[16].

In the construction domain, AutoML offers several advantages. Firstly, it simplifies the model development process by automating tasks requiring extensive manual effort, which enables construction professionals to focus more on domain-specific knowledge and problem-solving rather than the technical intricacies of machine learning[17]. Secondly, AutoML facilitates faster model development cycles, allowing construction projects to benefit from timely insights and predictions. AutoML expedites the overall model development process by automating repetitive tasks, making it more efficient and cost-effective[18]. Additionally, AutoML platforms often provide user-friendly interfaces and visualization tools that enable easier interpretation and communication of results to stakeholders in the construction industry.

While generative AutoML tools offer significant advantages, they are not without limitations. These include substantial computational resource requirements, time-intensive model generation and testing processes, and a lack of transparency that can complicate result interpretation and troubleshooting. The risk of overfitting is also a concern, as these tools may excessively learn from training data, leading to suboptimal performance on new data. Customization limitations may also arise, as users may occasionally wish to adjust a model in ways not supported by the tool. The effectiveness of generative AutoML tools is heavily reliant on the quality of input data; poor data quality can result in ineffective models. Finally, while some AutoML tools are freely available, others, particularly cloud-based platforms, can be costly.

## 3. Methodology

### 3.1 Research design

We utilized a comparative research design in this investigation, as illustrated in Figure 1. Our initial step involved the collection of representative industrial datasets from a variety of construction issues, which were of two types: classification and regression. Following this, we embarked on selecting typical AutoML methods, a process detailed in Section 3.3 and conducted via benchmark testing. In Section 3.4, we turned our attention to establishing evaluation criteria, providing a comprehensive discussion on how these performance metrics were determined. Lastly, we concluded our study by presenting the experimental results, offering valuable insights into the effectiveness of the methods and criteria used.

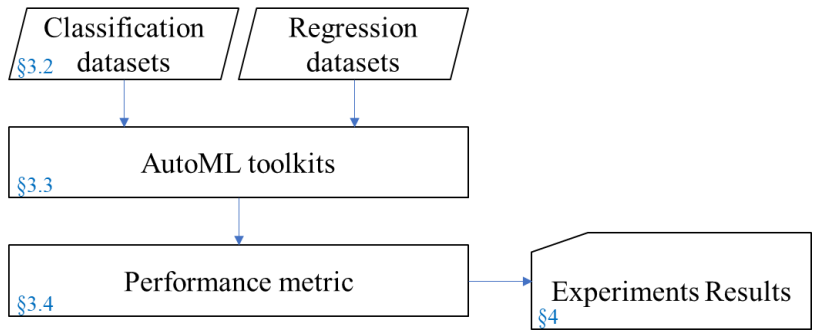

Figure 1. The research design

### 3.2 Datasets

Table II displays a collection of six datasets that pertain to the construction industry, comprised of both classification and regression datasets. These datasets present challenges from various sources, including construction projects, maintenance, materials, and real estate sales. The classification datasets contain discrete target values such as '0' or '1'[19], while the regression datasets contain continuous values like sales data[20]. The number of instances in these datasets ranges from 57 to 10,000. Notably, two classification datasets exhibit class imbalance, and two have limited instances, assuming a threshold of 1,000 instances. Additionally, these datasets' independent attributes range from 4 to 821. As indicated in the rightmost column of Table II, these datasets were originally published between 1998 and 2021.

    The classification datasets include the following: The first is the 'Construction Project Type Prediction' dataset presented by Yang, Xue [21], which consists of 2,451 instances and 821 attributes. The second, titled 'Illegal Construction Waste Dumping Actions', classifies waste dumping behavior in Hong Kong through statistical dumping behavior analysis [22]. The third classification dataset, presented by Matzka [23], reflects predictive maintenance issues encountered in the industry.

    As for the regression datasets, the first is the 'Concrete Compressive Strength Prediction' dataset, which represents a highly nonlinear function of age and composition[24]. The second regression dataset contains 9,568 data points collected from a cycle power plant over a span of six years. Lastly, Yeh and Hsu[25] collected a 'Residential Valuation' dataset of unit prices from Sindian District, New Taipei City, Taiwan.

Table II. List of collected construction datasets from the literature.

| Type | Id | Name | Target attribute | No. of instances | Imbalance? | Limited? | No. of attributes | Source |
|---|---|---|---|---|---|---|---|---|
| Classification | 1 | Construction project type prediction | "Building" or "Renovation" | 2,451 (1826; 625)* | Yes | No | 821 | (Yang et al. 2021) |
| Classification | 2 | Illegal construction waste dumping actions | "Yes" or "No" | 57 (29; 28) | No # | Yes | 54 | (Lu 2019) |
| Classification | 3 | AI4I 2020 predictive maintenance | "1" means successful or "0" means failed | 10,000 (9,661; 339) | Yes | No | 14 | (Matzka 2020) |
| Regression | 4 | Concrete compressive strength | Concrete compressive strength | 1,030 | - | No | 9 | (Yeh I. C. 1998) |
| Regression | 5 | Combined cycle power plant | Net hourly electrical energy | 9,568 | - | No | 4 | (Tüfekci 2014) |
| Regression | 6 | Real Estate Valuation | Residential unit price | 415 | - | Yes | 6 | (Yeh & Hsu 2018) |

*: (Number of instances of the majority class; number of instances of the minority class)
#: Manually balanced by the author.

### 3.3 Selection of AutoML tools

AutoML is increasingly accessible to industrial users. As listed in Table I, the AutoML tools selected for analysis fall into two categories: open-source and commercial. We chose Auto-sklearn (version 0.14.2) and H2O-AutoML (version 3.34.0) from the open-source category. From the commercial category, we have selected Azure Automated ML (version 1.37.0), Vertex AI (version 0.7), and EasyDL (version 2.0.12). It's important to note that the companies behind these commercial solutions, namely Microsoft, Google, and Baidu, are prominent technology giants, particularly recognized for their contributions to cloud computing and Artificial Intelligence.

### 3.4 Performance metric

This paper aims to assess various AutoML tools using two primary criteria: performance and costs. Performance is evaluated based on two key factors: error metrics (or accuracy) and compatibility with eight industrial datasets. For classification tasks, the primary metric is the macro F1 score, defined as the harmonic mean of precision and recall, which effectively reduces the impact of imbalanced datasets. For regression tasks, the root mean square error (RMSE) is used to compare the performance of regression models. The objective is to identify the model or machine learning pipeline with the lowest RMSE and highest F1 score. Error metrics are measured using 10-fold cross-validation for each dataset.

Regarding costs, AutoML tools are evaluated based on computational time, license fee, and ease of learning. Computational time only includes the standard platform execution time, excluding data formatting, cleansing, and network traffic. License fees are converted to US dollars. Ease of learning is assessed based on the availability of detailed documentation, clear error indication, and illustrative results presentation. The ultimate goal is to identify an AutoML tool that offers high performance at low costs.

## 4. Results and Discussion

To conduct a comprehensive comparison of open-source and commercial AutoML tools, we tested Auto-sklearn and H2O-AutoML on a personal computer equipped with an Intel(R) Core (TM) i7-7700 CPU (quad-core 3.60 GHz), 8 GB of memory, and a 64-bit Ubuntu (version 9.0.3). The maximum execution time for Auto-sklearn was set to 5 minutes, and we used all default parameters for H2O-AutoML version 3.34.0. In contrast, the commercial solutions were run on their respective cloud services. For the Azure experiments, we utilized the cost-effective 'Standard_D2s_v3' plan, which comes with 2 CPU cores, 8 GB of memory, and 16 GB of disk space. The hardware setting for Vertex AI experiments was not specified, and we optimized for 'value log loss'. EasyDL was configured with the default settings. It's important to note that most settings for commercial solutions are controlled by their respective cloud services providers, such as Microsoft and Google.

Table III compares the performance of the recommended AutoML tools with the results achieved by human experts in the literature across six datasets, excluding two datasets without academic publications. Generally, AutoML yielded more accurate results than those reported by human experts. Azure outperformed the literature in five out of six datasets. For instance, in the AI4I 2020 predictive maintenance for machines dataset, AutoML tools significantly improved the F1 score from Matzka's (2020) 0.79 to 0.803 (by H2O-AutoML), 0.865 (by Azure), and 0.885 (by Vertex AI). The only exception was the construction project type prediction, where Yang (2021) reported a satisfactory result with an F1 score of 0.873. The two recommended AutoML tools returned F1 scores of 0.855 and 0.843; however, Vertex AI found a slightly better result (F1 = 0.874). Azure surpassed the reported results in the literature in all the regression tasks. For example, Azure returned an R2 score of 0.94 for the concrete compressive strength dataset, which is better than the human expert's ML result of R2 = 0.86. The RMSE reported for the real estate valuation dataset is 7.73 in the literature, and the result with EasyDL is 5.75, indicating a 26% reduction in error in terms of RMSE.

Table III. Comparisons of accuracy in the literature and the two recommended AutoML tools (best values in bold)

| Test dataset | Performance* of ML | | |
|---|---|---|---|
|  | Reported in literature | Azure | H2O-AutoML |
| Construction project type prediction | $F_1 = 0.873$ | $F_1 = 0.855$ | $F_1 = 0.843$ |
| Illegal construction waste dumping | $F_1 = 0.860$ | $F_1 = 0.867$ | $F_1 = 0.854$ |
| AI4I 2020 predictive maintenance | $F_1 = 0.79$ | $F_1 = 0.865$ | $F_1 = 0.803$ |
| Concrete compressive strength | $R^2 = 0.86$ | $R^2 = 0.92$ | -[#] |
| Combined cycle power plant | RMSE = 3.78 | **RMSE = 3.48** | RMSE = 6.08 |
| Real estate valuation | RMSE = 7.73 | **RMSE = 6.73** | RMSE = 7.39 |

*: Higher is better for the first four datasets; lower is better for the last two.
#: Not reported by H2O-AutoML, but the RMSE was better than Azure's.

Table IV summarizes the costs associated with various AutoML tools as observed in our benchmarking evaluations. To create a more comprehensive ranking system, we devised five additional criteria across different categories to evaluate the cost-effectiveness of these tools. These criteria include:

- "In-house coding (20%)" indicates whether the AutoML practitioner is required to write computer codes.
- "Data type compatibility (20%)" assesses the tool's ability to handle different data types commonly found in the construction industry, thereby saving the practitioner's time from data conversion and preprocessing.
- "Minimum number of instances (15%)", which evaluates the tool's compatibility with different data volumes.
- "User friendliness (25%)" for which we invited new users with no programming skills to test the AutoML tools and collected their feedback on the software's user-friendliness.
- "Languages of user manuals (20%)" examines the range of languages available for the user manuals associated with each tool.

User friendliness is a key factor for teams with limited technical skills or for projects that require quick adoption of the tool. If ease of use is a priority, this criterion should be given more weight. If a tool is user-friendly, it can be more easily and quickly adopted by a team, especially if the team members have varying levels of technical expertise. On the other hand, the "Minimum number of instances" criterion is more specific to the technical capabilities of the tool and might not impact the user experience as directly.

According to the table, Azure achieved the highest overall score. Its exceptional efficiency, affordability, and extensive documentation make it an ideal choice for construction professionals lacking programming expertise. This high score is attributed to specific features and benefits of Azure that address the needs and challenges faced by construction practitioners.

Table IV. Overall comparison of the costs of AutoML tools used in this research (best in bold)

|  | Open-source | | Commercial | | |
| --- | --- | --- | --- | --- | --- |
|  | Auto-sklearn | H2O-AutoML | Azure | EasyDL | Vertex AI |
| Average computational time(min) | 5 | **1** | 37.5 | 18 | 157.17 |
| Total fee ($) | **0** | **0** | 70.08 | 31.44 | 255.02 |
| In-house coding | Yes | Yes | **No** | **No** | **No** |
| Data type compatibility (nominal and null values) | Numbers only | **OK** | **OK** | **OK** | **OK** |
| Minimum number of instances | **Any** | **Any** | **Any** | **Any** | 1,000 |
| User Friendliness* | Hard to use for beginners | No messages for explaining errors | **Yes** | Unfriendly guideline | **Yes** |
| Languages of user manuals | English | English | **Multi-language** | Chinese | **Multi-language** |
| Wins | 2 | 4 | **5** | 3 | 4 |

*: rated by five fresh users.

## 5. An application case of construction project type prediction

Step 1 consists of two parts, namely data collection and AutoML preparation. The example case's data collection, including cleansing and attribute definitions, was reported in detail by Yang et al. (2021). Figure 6 shows the AutoML preparation on Azure. Figure 6(a) shows the definition of a new dataset. Figure 6(b) shows the "datastore and file selection" dialog, where one can upload the collected dataset. The "Schema" dialog in Figure 6(c) offers a data filter to exclude unnecessary attributes and labels, such as the unique identification columns or alternative target classes. Figure 6(d) shows the creation of an AutoML job, which automatically pops up after the data filtering. Figure 6(e) is the configuration dialog of computational resources for the defined AutoML job. We recommend the low-cost virtual machine "Standard_DS11_v2" for the small to medium sized datasets collected from the construction industry. If the dataset contains a large number (e.g., > 1 million) of instances, advanced virtual machines are recommended.

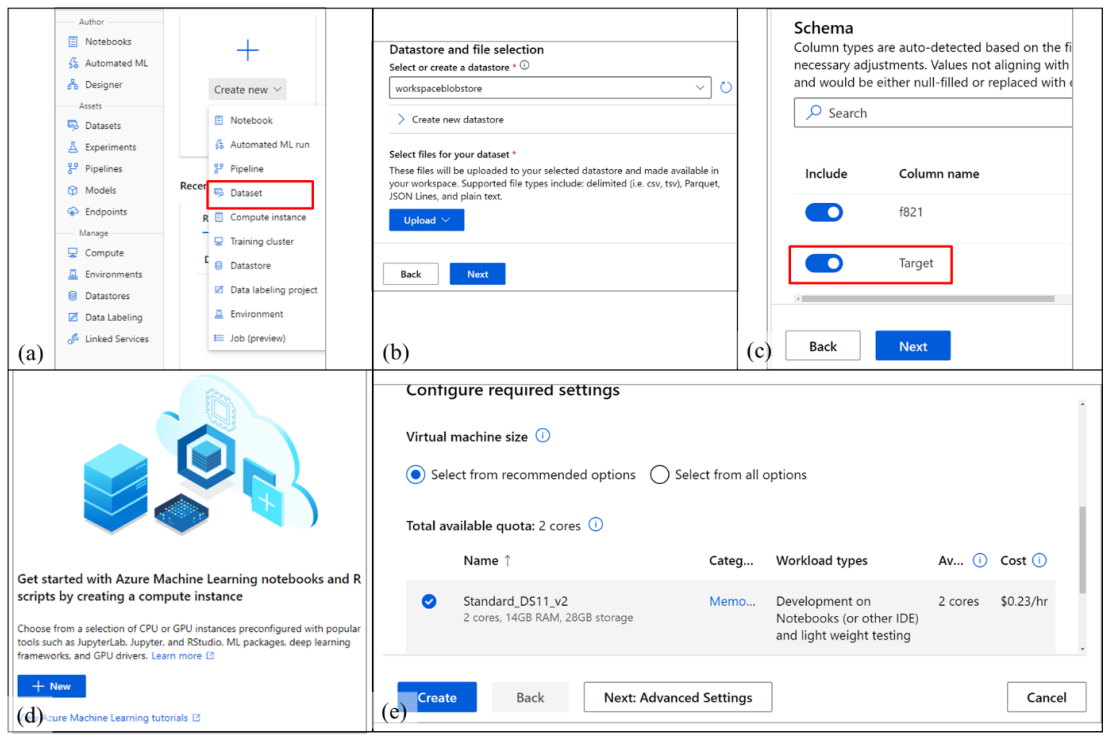

Figure 6. Demonstration of Step 1 of the AutoML guideline

In Step 2, AutoML trains (or re-trains) candidate ML models. Figure 7(a) shows that Azure offers a "+ New Automation ML Run" button to start the training. The "select data" dialog in Figure 7(b) lists the available datasets to work with. Figure 7(c) shows the "Configure Run" dialog to create a new AutoML job. Then, one can follow "classification" for discrete target classes (e.g., types of construction projects) or "regression" for continuous targets (e.g., property sales) in the job settings in Figure 7(d) to start the AutoML.

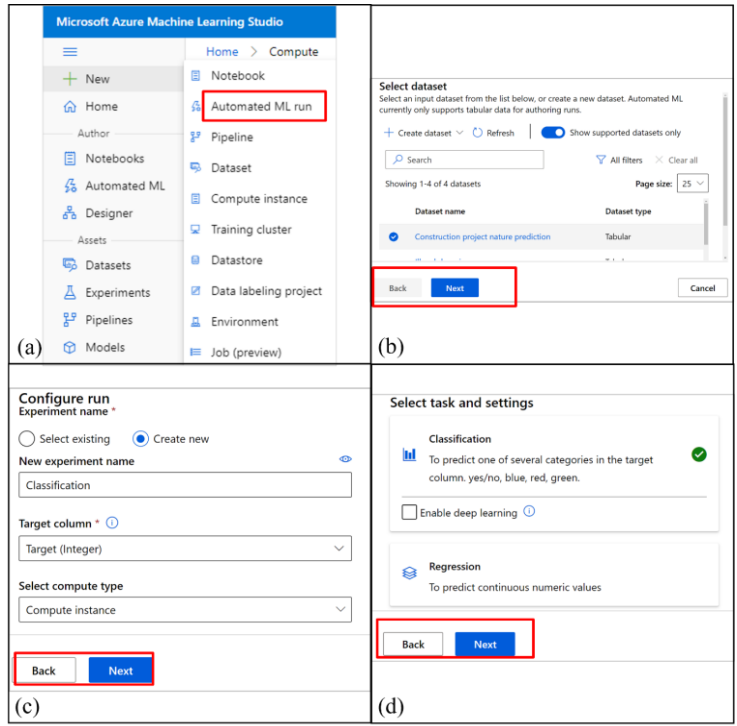

Figure 7. Demonstration of Step 2 of the AutoML guideline

In step 3, one can evaluate the classification or regression results trained by AutoML. When the running status on the dialog in Figure 8(a) changes to "Completed," the AutoML job is completed in Azure. Figure 8(b) shows the trained ML models with metric scores under the "Models" tab. In addition, the web links in the model list explain the associated ML model and patterns. Advanced properties, metrics, and performance charts are available in the "Metrics" tab, as shown in Figure 8(c). Although there are no general thresholds of acceptance for classification and regression, a higher balanced accuracy metric such as "macro F1 score" or a lower error metric such as RMSE is always preferred. If the resulting metric is unsatisfactory, e.g., $F_1 < 0.65$ in classifying the water portability dataset, one can rewind Step 2 or Step 1 to revise the AutoML job settings or dataset.

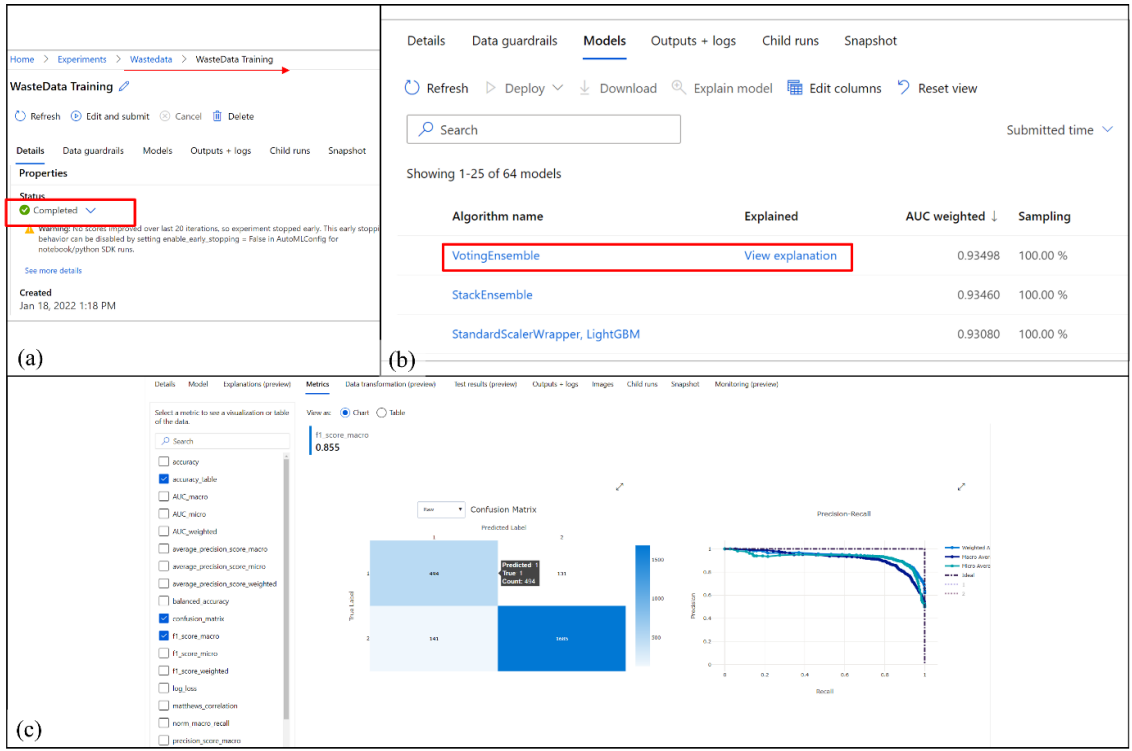

Figure 8. Demonstration of Step 3 of the AutoML guideline

Some AutoML tools also offer advanced metrics and charts for evaluating the input dataset and training results. For example, Figure 9(a) shows the top 4 important attributes in the dataset with their important values. One should revise the definitions or add more attributes when no important ones are discovered. Figure 9(b) shows a trained ML model's ensemble weights and hyperparameters in JSON format. Figure 9(c) depicts the data processing by AutoML in a directed acyclic graph, in which the data preprocessing, feature selection, scaling strategies, and ML models are optimized by AutoML. According to the data metrics and charts, a construction practitioner can review and revise the dataset, e.g., updating the definitions of attributes, correcting faults, handling data missingness, and supplementing extra instances.

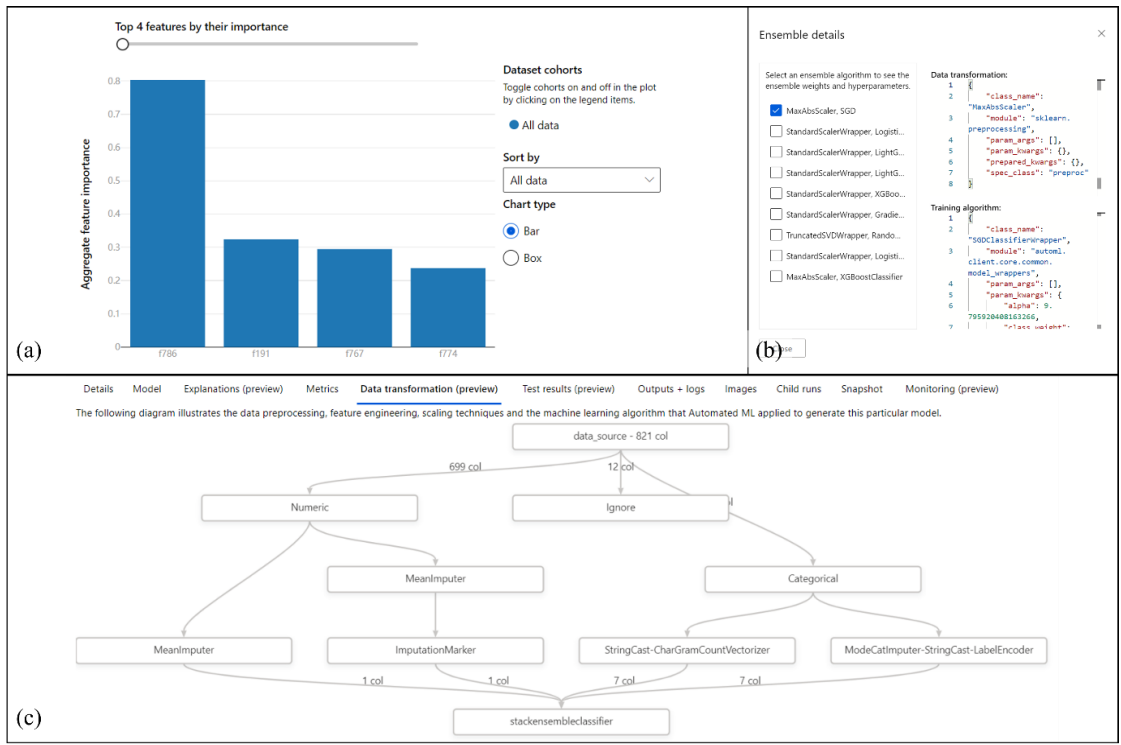

Figure 9. Detail parameters of training model

In Step 4, the ML model trained by AutoML and confirmed by the construction practitioner is applied to the target industrial scenario. For commercial AutoML tools such as Azure, the trained model is usually available as an online cloud service, as shown in Figures 10(a) and 10(b). After the successful deployment as a cloud service, the model trained by AutoML can be accessed through Power BI. Figure 10(c) shows the Power BI dialog for importing a data file (e.g., in Excel format) from local disks. The classification or prediction results are available on the dialog, as shown in Figure 10(d), under the "Azure Machine Learning" option. Alternatively, the model trained by AutoML can also be deployed as a "real-time endpoint," as shown in Figure 10(a), when access to the Internet is limited.

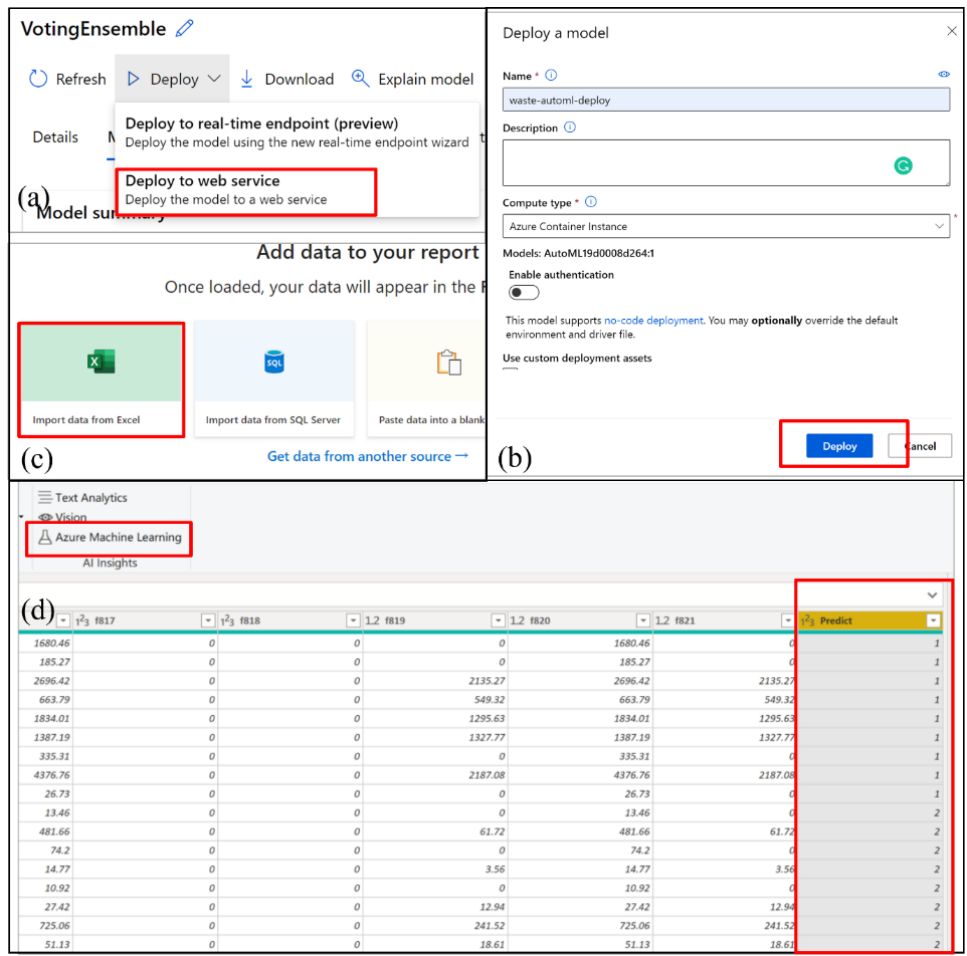

Figure 10. Demonstration of Step 4 of the AutoML guideline

## 6. Conclusion

Smart construction is expected to improve the construction industry's efficiency, safety, and sustainability by applying digital technology and advanced building practices. The findings in this paper successfully demonstrated the potential of AutoML is revolutionizing the construction industry. The results from six classification and regression tasks of various datasets have shown that AutoML can simplify the traditionally complex process of constructing predictive models, making it accessible to professionals without extensive data science training. Based on the findings, we encourage construction practitioners to embrace the novel AutoML techniques and continue exploring their potential to transform construction practices. In selecting an AutoML tool, we first need to consider user-friendliness and ease of use. The cost of use is also an essential aspect, and as technology develops, the accuracy and time spent on models will improve.

However, it is important to acknowledge the limitations of this paper. While we have shown the potential of AutoML in a specific case study, further research is needed to explore its applicability to a wider range of construction scenarios and datasets. Additionally, the challenges unique to construction datasets, such as their complexity and variability, require further investigation to optimize the use of AutoML in this industry fully. Future research directions are suggested to explore the combination of machine learning and the latest technologies, such as blockchain, virtual reality, and digital twins.

## Acknowledgment

The work presented in this paper was supported by Hong Kong Research Grants Council (Nos. T22-504/21-R and C7080-22GF) and in part by Shenzhen Science, Technology and Innovation Commission (No. SGDX20201103093600002).